\documentclass[aps, prb, amsmath, amssymb, english, twocolumn,reprint,floatfix, superscriptaddress]{revtex4-2}

\usepackage[english]{babel}
\usepackage{graphicx}
\usepackage{verbatim}

\usepackage{epsfig}
\usepackage{epstopdf}
\usepackage{amsfonts}
\usepackage{amsthm}
\usepackage{amsmath}
\usepackage{amssymb}
\usepackage{color}
\usepackage[usenames,dvipsnames,svgnames,table]{xcolor}
\usepackage{hyperref} 
\hypersetup{pdfpagemode=FullScreen,colorlinks=true,linkcolor=NavyBlue,citecolor=BrickRed}
\usepackage{makeidx}
\usepackage{pifont}

\usepackage{booktabs}
\usepackage{xr}
\usepackage{color}
\usepackage{grffile}

\usepackage[ruled]{algorithm2e}
\usepackage[noend]{algpseudocode}
\usepackage{multirow}
\usepackage{type1cm} 
\usepackage{dcolumn}
\usepackage{bm}

\begin{document}

\title{Orbital-dependent electronic correlations\\and single-band character in layered palladates}
\author{Jing-Xuan Wang}
\affiliation{School of Physics and Key Laboratory of Quantum State Construction and Manipulation (Ministry of Education), Renmin University of China, Beijing 100872, China}
\author{Jia-Ming Wang}
\affiliation{Center for Materials Theory, Department of Physics and Astronomy, Rutgers University, Piscataway, New Jersey 08854, USA}
\author{Rong-Qiang He}\email{rqhe@ruc.edu.cn}
\affiliation{School of Physics and Key Laboratory of Quantum State Construction and Manipulation (Ministry of Education), Renmin University of China, Beijing 100872, China}
\author{Zhong-Yi Lu}\email{zlu@ruc.edu.cn}
\affiliation{School of Physics and Key Laboratory of Quantum State Construction and Manipulation (Ministry of Education), Renmin University of China, Beijing 100872, China}
\affiliation{Hefei National Laboratory, Hefei 230088, China}

\date{\today}

\begin{abstract}

Layered palladates provide a $4d$ platform for examining electronic correlations in materials with cuprate-like structural and orbital characteristics. Using fully charge-self-consistent density functional theory plus dynamical mean-field theory, we investigate the orbital- and layer-resolved electronic structures of LaPdO$_2$, RbSr$_2$PdO$_3$, BaLaPdO$_2$Cl$_2$, La$_4$Pd$_3$O$_8$, and La$_5$Pd$_4$O$_{10}$. Under the conditions studied, all compounds remain metallic, with low-energy states dominated by the Pd-$d_{x^2-y^2}$ orbital. This orbital exhibits stronger mass renormalization than the Pd-$d_{z^2}$ orbital. LaPdO$_2$ and BaLaPdO$_2$Cl$_2$ show moderate correlations, whereas RbSr$_2$PdO$_3$ is more weakly correlated. Both the trilayer La$_4$Pd$_3$O$_8$ and quadlayer La$_5$Pd$_4$O$_{10}$ remain weakly correlated, with similar Pd-$d_{x^2-y^2}$ mass enhancements and little difference between inner- and outer-layer Pd sites. Among the structures studied, BaLaPdO$_2$Cl$_2$ is closest to a cuprate-like single-band description, whereas RbSr$_2$PdO$_3$ retains an additional electron pocket with Pd-$d_{z^2}$ character. These results establish a comparative picture of weak to moderate, orbital-dependent electronic correlations in layered palladates.


\end{abstract}


\maketitle

\section{INTRODUCTION}
Since the discovery of high-temperature superconductivity in layered cuprates, identifying materials with analogous low-energy electronic structures has remained a central theme in the search for new unconventional superconductors~\cite{Bednorz1986}. Cuprate superconductors are characterized by CuO$_2$ planes, a nominal Cu $d^9$ configuration, and a strongly correlated low-energy electronic structure dominated near the Fermi level by a single antibonding band of primarily Cu $d_{x^2-y^2}$ and O $p_\sigma$ character. Infinite-layer nickelates, $R$NiO$_2$ ($R$ = rare earth), provide an important extension because they contain square-planar NiO$_2$ layers with nominal Ni $d^9$ filling~\cite{Anisimov1999,Lee2004,xiao2024superconductivity,li2019superconductivity}. The parent infinite-layer nickelates show metallic or bad-metal-like behavior, rather than canonical charge-transfer insulating behavior, due in part to self-doping associated with rare-earth $5d$-derived electron pockets~\cite{Lee2004,hepting2020electronic}. Nevertheless, the correlated Ni $3d_{x^2-y^2}$ states remain a dominant component of the low-energy electronic structure. This example motivates the exploration of other transition-metal oxides with tunable orbital character, filling, and correlation strength.

Layered palladates provide another platform for exploring this design principle. Similar to cuprates and nickelates, several palladates contain square-planar PdO$_2$ layers and formal Pd $d$-electron counts close to $d^9$ or slightly hole-doped $d^9$, making them promising candidates for realizing $d_{x^2-y^2}$-dominated low-energy states. A key difference is that Pd $4d$ orbitals are more spatially extended than Cu $3d$ and Ni $3d$ orbitals. The greater spatial extent of Pd-$4d$ orbitals generally favors larger hopping amplitudes and broader bands. Consequently, palladates are generally expected to have a smaller effective interaction-to-bandwidth ratio $U/W$ and weaker electronic correlations than cuprates. However, square-planar PdO$_2$ layers and near-$d^9$ filling alone do not establish a cuprate-like low-energy electronic structure. The strength of local correlations and the contributions of orbitals other than Pd-$d_{x^2-y^2}$ near the Fermi level must also be examined. 


Superconductivity has been reported in several Pd-containing compounds, including Li$_2$Pd$_3$B, SrPd$_2$Ge$_2$, PdTe$_2$, and Ta$_4$Pd$_3$Te$_{16}$~\cite{Badica2006,Kim2012prb,jiao2014superconductivity,gupta2024prb}. However, these materials do not realize the square-planar PdO$_2$ layers or the cuprate-like Pd $d_{x^2-y^2}$ low-energy states relevant to layered palladium oxides. Previous theoretical work used first-principles and multiorbital DMFT calculations to examine the electronic structures of layered palladates and investigated superconductivity within an effective single-band Hubbard model. Within this model, intermediate coupling, moderate Fermi-surface warping, and low hole doping were found to favor higher superconducting transition temperatures. Based on these results, RbSr$_2$PdO$_3$ and $A'_2$PdO$_2$Cl$_2$ ($A'$ = Ba$_{0.5}$La$_{0.5}$) were proposed as nearly optimal palladate candidates, whereas compounds such as NdPdO$_2$ were found to be too weakly correlated~\cite{Kitatani2023prl}. A complementary ab initio study of layered palladium oxides showed that LaPdO$_2$, despite its nominal $d^9$ configuration, differs substantially from cuprates, whereas La$_4$Pd$_3$O$_8$ has an electronic structure closer to that of overdoped cuprates~\cite{botana2018prm}. Kitatani \textit{et al.} also reported small additional pockets in the PBE bands of RbSr$_2$PdO$_3$ and showed that hole doping or a modified Becke--Johnson potential can remove the zone-center pocket~\cite{Kitatani2023prl}. A central question is how the strength and orbital dependence of local correlations vary across these layered palladates and how they relate to the single-band character of the low-energy states. These studies motivate a comparison of orbital- and layer-resolved correlations across the proposed palladates within a charge-self-consistent DFT+DMFT framework.


This question has become particularly timely in light of the recent observation of superconducting signatures in multi-layer square-planar nickelates Nd$_{n+1}$Ni$_n$O$_{2n+2}$ for $n=4$--8, where the carrier concentration and dimensionality can be tuned by the number of NiO$_2$ layers without explicit chemical doping~\cite{pan2026superconducting}. In this series, the $n=4$ compound Nd$_5$Ni$_4$O$_{10}$ has an average Ni valence of $+1.25$, corresponding to a nominal $d^{8.75}$ configuration. This suggests that finite-layer square-planar oxides with slightly hole-doped $d^9$ filling provide a useful route for searching for cuprate-related superconducting platforms. First-principles many-body calculations for Pr$_4$Ni$_3$O$_8$ and Nd$_6$Ni$_5$O$_{12}$ have further highlighted the interplay of strongly correlated Ni-$d_{x^2-y^2}$ states and low-energy Ni-$d_{z^2}$ states~\cite{lechermann2022emergent}. In this context, La$_5$Pd$_4$O$_{10}$ represents a natural four-layer palladate analogue with the same nominal $d^{8.75}$ filling, while La$_4$Pd$_3$O$_8$ provides a closely related trilayer reference that was previously proposed to have an overdoped-cuprate-like electronic structure~\cite{botana2018prm}. Recent ab initio work on the La$_{n+1}$Pd$_n$O$_{2n+2}$
series with $n=4$--7 reported broader bands, stronger $p$--$d$ hybridization, and reduced rare-earth-derived contributions at the Fermi level compared with the corresponding nickelates~\cite{gavrilov2026electronic}. These findings motivate an examination of orbital- and layer-dependent dynamical correlations in La$_4$Pd$_3$O$_8$ and La$_5$Pd$_4$O$_{10}$ and their effects on the low-energy electronic structure.

In this work, we use fully charge-self-consistent DFT+DMFT calculations to compare the correlated electronic structures of LaPdO$_2$, RbSr$_2$PdO$_3$, BaLaPdO$_2$Cl$_2$, La$_4$Pd$_3$O$_8$, and La$_5$Pd$_4$O$_{10}$. By choosing compounds with different local distortions, spacer layers, nominal fillings, and numbers of PdO$_2$ layers, we examine the conditions under which a Pd-$d_{x^2-y^2}$-dominated cuprate-like low-energy structure can be realized in palladates. We focus on the orbital dependence of local electronic correlations and its relation to the low-energy electronic structure. In particular, we compare the renormalization of the Pd $d_{x^2-y^2}$ and $d_{z^2}$ states, assess the single-band character of the low-energy bands, and examine the layer dependence of local correlations in the multilayer compounds.


\section{METHODS}
In this study, we performed fully charge-self-consistent density functional theory plus dynamical mean-field theory (DFT+DMFT) calculations. Structural relaxations and preliminary electronic-structure calculations were performed using the Vienna \textit{ab initio} Simulation Package (VASP)~\cite{kresse1994norm,kresse1996efficiency}. The exchange-correlation functional was treated within the generalized-gradient approximation (GGA) in the Perdew-Burke-Ernzerhof (PBE) form~\cite{perdew1996generalized}. The optimized crystal structures were then used as input for the subsequent DFT+DMFT calculations. The fully charge self-consistent DFT+DMFT calculations were carried out using the eDMFT package~\cite{haule2010dynamical,haule2015free}, implemented on top of the full-potential linearized augmented-plane-wave code WIEN2k~\cite{Blaha2020jcp}.

For the main calculations, we used the same local interaction parameters, $U=5~\mathrm{eV}$ and $J_H=1~\mathrm{eV}$, for the full five-orbital Pd-$4d$ shell in all five palladates. These parameters refer to localized Pd-$4d$ orbitals constructed by projection within an energy window from $-10$ to $10$ eV relative to the Fermi level. Effective local interactions depend on the orbital representation and the screening channels associated with the chosen correlated subspace, and therefore are not directly transferable between different low-energy models~\cite{PhysRevB.86.165105}. Comparable interaction scales, $U=4.5~\mathrm{eV}$ and $J_H=1~\mathrm{eV}$, have been employed in wide-window DFT+DMFT calculations for $4d$ ruthenates~\cite{PhysRevLett.116.256401,deng2019signatures}. The sensitivity to $U$ is examined in Appendix~\ref{appD:U} through additional five-orbital Pd-$4d$ calculations for BaLaPdO$_2$Cl$_2$ at fixed $J_H=1~\mathrm{eV}$.

All DFT+DMFT calculations were performed at $\beta=40~\mathrm{eV}^{-1}$, corresponding to $T\approx290$ K. The main-text results were obtained using structures optimized at 0 GPa for LaPdO$_2$ and RbSr$_2$PdO$_3$, 5 GPa for BaLaPdO$_2$Cl$_2$, and 20 GPa for La$_4$Pd$_3$O$_8$ and La$_5$Pd$_4$O$_{10}$. For the latter three compounds, elevated pressures were used because their high-symmetry reference structures exhibit phonon instabilities at ambient pressure. A systematic search for dynamically stable lower-symmetry structures of these compounds at ambient pressure is beyond the scope of the present work. The phonon calculations in Appendix~\ref{app:structure} refer to these selected structures.

The impurity problem was solved using the continuous-time quantum Monte Carlo (CT-QMC) impurity solver in the hybridization-expansion formulation~\cite{gull2011continuous}. We used the exact double-counting scheme~\cite{haule2015exact}. After convergence on the Matsubara frequency axis, the real-frequency self-energy and spectral functions were obtained by analytic continuation using the maximum entropy method~\cite{jarrell1996bayesian}. The quasiparticle mass enhancement was estimated from the low-frequency slope of the Matsubara self-energy,
\begin{equation}
\frac{m^*}{m} = \frac{1}{Z}
\simeq 1-\left.\frac{\partial \operatorname{Im}\Sigma(i\omega_n)}
{\partial \omega_n}\right|_{\omega_n\rightarrow 0}.
\end{equation}
In practice, the lowest four positive Matsubara-frequency points of $\mathrm{Im}\Sigma(i\omega_n)$ were fitted to a quadratic form with a free intercept,
\begin{equation}
\operatorname{Im}\Sigma(i\omega_n)=a\omega_n^2+b\omega_n+c.
\label{eq:self-energy-fit}
\end{equation}
Here, $c$ is the fitted intercept, and the linear coefficient $b$ gives $Z^{-1}=1-b$. We considered only paramagnetic solutions and neglected spin-orbit coupling effects. Additional numerical details are provided in Appendix~\ref{appA:method}.

\section{RESULTS AND DISCUSSION}

\begin{figure}[htb]
  \centering
  \includegraphics[width=8.6cm]{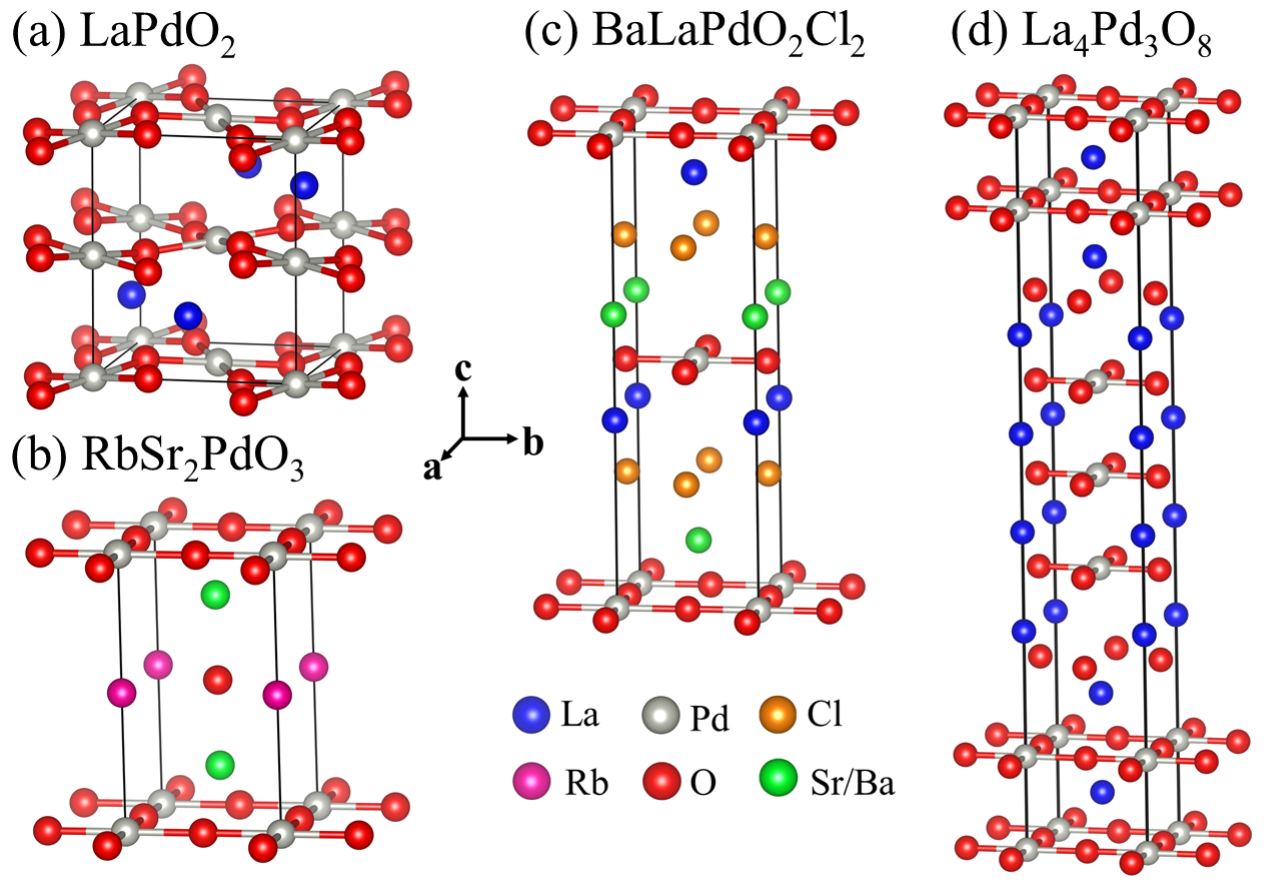}
  \caption{Crystal structures of representative layered palladates (a) LaPdO$_2$, (b) RbSr$_2$PdO$_3$, (c) BaLaPdO$_2$Cl$_2$, and (d) La$_4$Pd$_3$O$_8$.}
  \label{fig:structure}
\end{figure}

\subsection{Crystal structures}

Fig.~\ref{fig:structure} shows the crystal structures of
LaPdO$_2$, RbSr$_2$PdO$_3$, BaLaPdO$_2$Cl$_2$, and
La$_4$Pd$_3$O$_8$. The present study also includes La$_5$Pd$_4$O$_{10}$ as a four-layer counterpart of La$_4$Pd$_3$O$_8$. All five compounds contain PdO$_2$ layers with approximately square-planar PdO$_4$ units, but differ in local symmetry, spacer-layer environment, layer number, and nominal Pd-$d$ filling.

LaPdO$_2$ serves as an infinite-layer-like palladate reference with a formal Pd$^{1+}$, $4d^9$ configuration. The structure considered here has the \textit{Pnma} space group and tilted PdO$_4$ units, distinguishing it from an ideal tetragonal infinite-layer structure. This distortion can modify the local crystal field and the relative energies of the Pd-$d_{x^2-y^2}$ and Pd-$d_{z^2}$ states. RbSr$_2$PdO$_3$ and BaLaPdO$_2$Cl$_2$ also have a formal $4d^9$ configuration, but contain additional spacer layers between the PdO$_2$ planes. The Rb/Sr--O layers in RbSr$_2$PdO$_3$ and the Cl-containing blocking layers in BaLaPdO$_2$Cl$_2$
provide distinct interlayer environments that can affect the Pd orbital energies and interlayer-state admixture near the Fermi level.

La$_4$Pd$_3$O$_8$ and La$_5$Pd$_4$O$_{10}$ contain three and four PdO$_2$ layers per structural block, respectively. La$_4$Pd$_3$O$_8$ has an average formal Pd configuration of approximately $4d^{8.67}$, while La$_5$Pd$_4$O$_{10}$ has an average formal configuration of $4d^{8.75}$. Both therefore have finite nominal hole counts relative to $4d^9$. Their comparison allows us to examine the combined effects of nominal filling and interlayer coupling on the low-energy electronic structure.

For the optimized structures used in the main-text calculations, no imaginary phonon frequencies were found along the calculated high-symmetry paths. The optimized lattice parameters and phonon dispersions are provided in Appendix~\ref{app:structure}. The pressure dependence of the Pd-$d_{x^2-y^2}$ self-energies in the multilayer compounds is discussed in Appendix~\ref{appC:pressure}.

\begin{table}[t]
    \renewcommand\arraystretch{1.5}
    \caption{Local Pd-$4d$ orbital occupations $N_d$ and effective mass enhancements $m^*/m=Z^{-1}$ for the materials studied. The labels (i) and (o) denote inner- and outer-layer Pd sites, respectively.}\label{tab1}
    \label{tab:occupancy}
    \begin{ruledtabular}
    \begin{tabular}{lccccccc}
    & \multicolumn{5}{c}{$N_d$} & \multicolumn{2}{c}{$m^*/m$} \\ \cline{2-6} \cline{7-8}
    Material & $d_{z^2}$ & $d_{x^2-y^2}$ & $d_{xz/yz}$ & $d_{xy}$ &$d$& $d_{x^2-y^2}$ & $d_{z^2}$ \\
    \hline
    LaPdO$_2$          & 1.73 & 1.33 & 1.90 & 1.91 &8.77 & 2.03 & 1.14 \\
    RbSr$_2$PdO$_3$   & 1.64 & 1.31 & 1.91 & 1.94 &8.71  & 1.56 & 1.12 \\
    BaLaPdO$_2$Cl$_2$ & 1.73 & 1.32 & 1.91 & 1.93 & 8.80 & 1.71 & 1.12 \\
    La$_4$Pd$_3$O$_8$ (i) & 1.67 & 1.16 & 1.90 & 1.94 &8.57  & 1.50 & 1.11 \\
    La$_4$Pd$_3$O$_8$ (o) & 1.68 & 1.14 & 1.91 & 1.93 &8.57  & 1.51 & 1.10 \\
    La$_5$Pd$_4$O$_{10}$ (i) & 1.66 & 1.21 & 1.90 & 1.93& 8.60 & 1.53 & 1.11 \\
    La$_5$Pd$_4$O$_{10}$ (o) & 1.68 & 1.17 & 1.91 & 1.92&8.59  & 1.54 & 1.10 \\
    \end{tabular}
    \end{ruledtabular}
\end{table}

\begin{figure*}[htb]
  \centering
  \includegraphics[width=18cm]{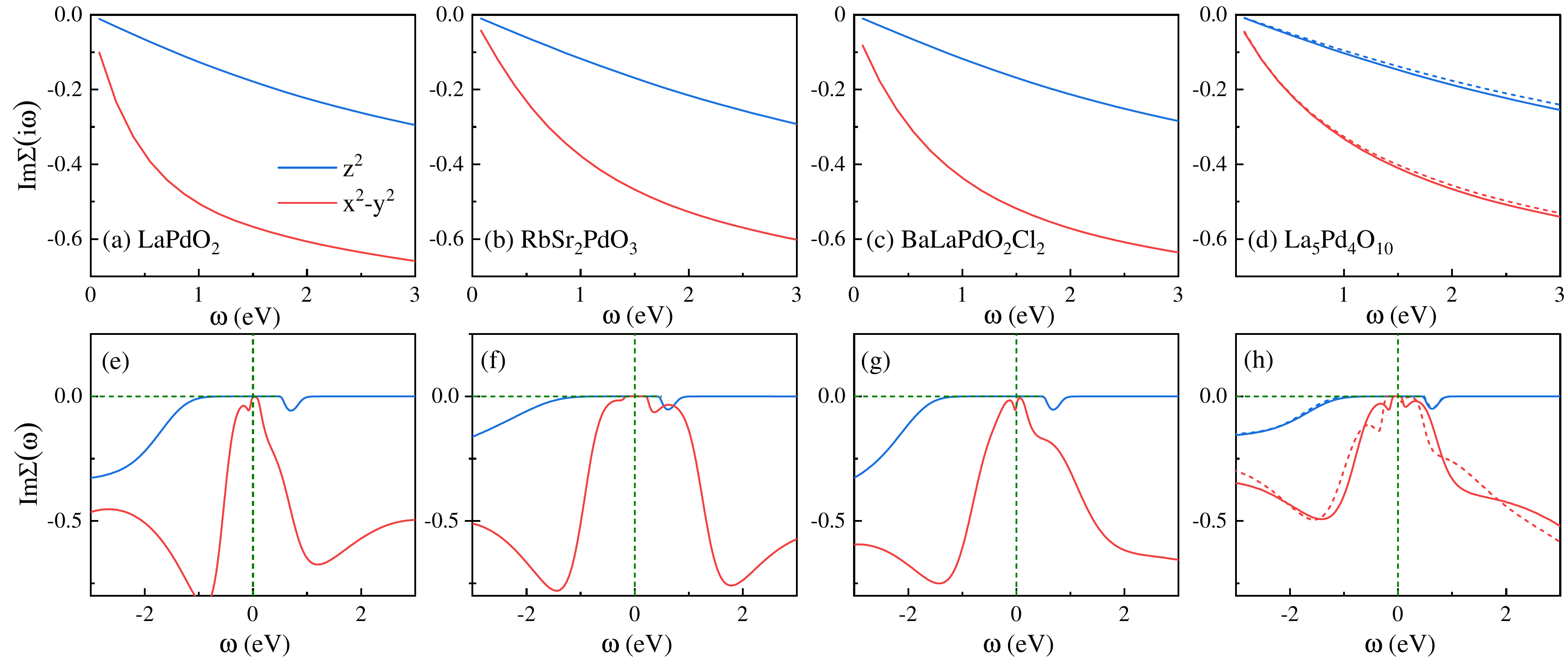}
  \caption{Orbital-resolved self-energies of the four representative layered palladates obtained from DFT+DMFT at $\beta=40$ eV$^{-1}$, i.e., 290 K. Panels (a)--(d) show the imaginary parts of the Matsubara-frequency self-energies, $\mathrm{Im}\Sigma(i\omega_n)$, for LaPdO$_2$, RbSr$_2$PdO$_3$, BaLaPdO$_2$Cl$_2$, and La$_5$Pd$_4$O$_{10}$, respectively. Panels (e)--(h) show the corresponding imaginary parts of the real-frequency self-energies, $\mathrm{Im}\Sigma(\omega)$. Blue and red curves correspond to the Pd-$d_{z^2}$ and Pd-$d_{x^2-y^2}$ orbitals, respectively. In panels (d) and (h), solid and dashed curves denote the inner- and outer-layer Pd sites, respectively. The calculations use structures optimized at 0 GPa for LaPdO$_2$ and RbSr$_2$PdO$_3$, 5 GPa for BaLaPdO$_2$Cl$_2$, and 20 GPa for La$_5$Pd$_4$O$_{10}$.}
  \label{fig:self-energy}
\end{figure*}

\subsection{Local occupancies and mass enhancement}

Table~\ref{tab1} summarizes the local Pd-$4d$ orbital occupations $N_d$ and effective mass enhancements $m^*/m=Z^{-1}$ obtained from the DFT+DMFT calculations at 290~K. The $t_{2g}$ orbitals are nearly fully occupied with occupations of 1.90--1.94 per orbital. The Pd $d_{z^2}$ orbital has occupations ranging from 1.64 to 1.73 and exhibits weak mass renormalization with $Z^{-1}=1.10$--1.14.

LaPdO$_2$, RbSr$_2$PdO$_3$, and BaLaPdO$_2$Cl$_2$ have similar Pd-$d_{x^2-y^2}$ occupations of 1.33, 1.31, and 1.32, respectively, but their mass enhancements differ appreciably. The Pd-$d_{x^2-y^2}$ mass enhancements $Z^{-1}$ are 2.03 in LaPdO$_2$ and 1.71 in BaLaPdO$_2$Cl$_2$, indicating moderate correlations in this orbital. In RbSr$_2$PdO$_3$, the Pd-$d_{x^2-y^2}$ mass enhancement is 1.56, indicating weaker correlations than in LaPdO$_2$ and BaLaPdO$_2$Cl$_2$.

For La$_4$Pd$_3$O$_8$ and La$_5$Pd$_4$O$_{10}$, the crystallographically inequivalent Pd sites were treated as separate impurity problems in the DFT+DMFT calculations. 
As shown in Table~\ref{tab1}, the inner- and outer-layer Pd sites exhibit very similar orbital occupations, indicating weak layer dependence. The Pd-$d_{x^2-y^2}$ occupations are slightly higher in La$_5$Pd$_4$O$_{10}$ (1.17--1.21) than in La$_4$Pd$_3$O$_8$ (1.14--1.16). Both multilayer compounds have lower Pd-$d_{x^2-y^2}$ occupations than the three nominal $4d^9$ compounds, for which the values are 1.31--1.33. This trend is consistent with the nominal hole counts relative to $4d^9$, which are zero for the three $4d^9$ compounds, 0.25 for La$_5$Pd$_4$O$_{10}$, and 0.33 for La$_4$Pd$_3$O$_8$. Although the nominal hole count decreases from the trilayer to the quadlayer, the Pd-$d_{x^2-y^2}$ mass enhancements remain similar at $Z^{-1}=1.50$--1.51 and 1.53--1.54, respectively, suggesting weak correlations in both compounds.

For the corresponding square-planar nickelates, She \textit{et al.} reported Ni-$d_{x^2-y^2}$ occupations of 1.003--1.018 in La$_4$Ni$_3$O$_8$ and 1.013--1.033 in La$_5$Ni$_4$O$_{10}$~\cite{she2026layer}. The Ni-$d_{x^2-y^2}$ occupations are closer to one electron per orbital than the Pd occupations calculated here at 20 GPa.
The reported nickelate occupations are slightly higher in the quadlayer than in the trilayer.



Compared with typical Cu-$3d$-based cuprates, the Pd-$4d$ orbitals are spatially more extended, which generally leads to a larger effective bandwidth and a reduced local correlation strength. Therefore, although these palladates contain a cuprate-like Pd-$d_{x^2-y^2}$-derived low-energy channel, their electronic correlations remain weak to moderate within the present DFT+DMFT calculations. The calculated spectra remain metallic at 290 K.

\subsection{Self-energies}


To quantify the correlation effects, the upper panels (a)--(d) of Fig.~\ref{fig:self-energy} display the imaginary parts of the Matsubara self-energies, $\mathrm{Im}\Sigma(i\omega_n)$, for LaPdO$_2$, RbSr$_2$PdO$_3$, BaLaPdO$_2$Cl$_2$, and La$_5$Pd$_4$O$_{10}$, respectively. The lower panels (e)--(h) show the corresponding $\mathrm{Im}\Sigma(\omega)$ on the real-frequency axis. For all four compounds, the magnitude of $\mathrm{Im}\Sigma(i\omega_n)$ for the Pd $d_{x^2-y^2}$ orbital is substantially larger than that for the Pd $d_{z^2}$ orbital, indicating that the correlation effects are mainly concentrated in the in-plane $d_{x^2-y^2}$ orbital. In contrast, the Pd $d_{z^2}$ orbital exhibits a much weaker and smoother low-frequency self-energy on the Matsubara axis, and its real-frequency $\mathrm{Im}\Sigma(\omega)$ also remains close to zero near $\omega=0$. This behavior indicates weaker correlation effects in the Pd-$d_{z^2}$ orbital, consistent with Table~\ref{tab1}.

For the Pd $d_{x^2-y^2}$ orbital, the Matsubara self-energies $\mathrm{Im}\Sigma(i\omega_n)$ in LaPdO$_2$ and BaLaPdO$_2$Cl$_2$ show a steeper low-frequency slope together with an apparent finite low-frequency intercept. The corresponding imaginary parts of the real-frequency self-energies also remain finite near $\omega=0$, indicating stronger low-energy scattering and a more strongly renormalized metallic state at 290 K. By contrast, the Pd $d_{x^2-y^2}$ self-energies in RbSr$_2$PdO$_3$ and La$_5$Pd$_4$O$_{10}$ exhibit a linear low-frequency dependence on the Matsubara axis and the corresponding imaginary parts of the real-frequency self-energies remain close to zero near $\omega=0$. This behavior indicates more coherent quasiparticles and is consistent with Fermi-liquid behavior.

\begin{figure}[htb]
  \centering
  \includegraphics[width=8.6cm]{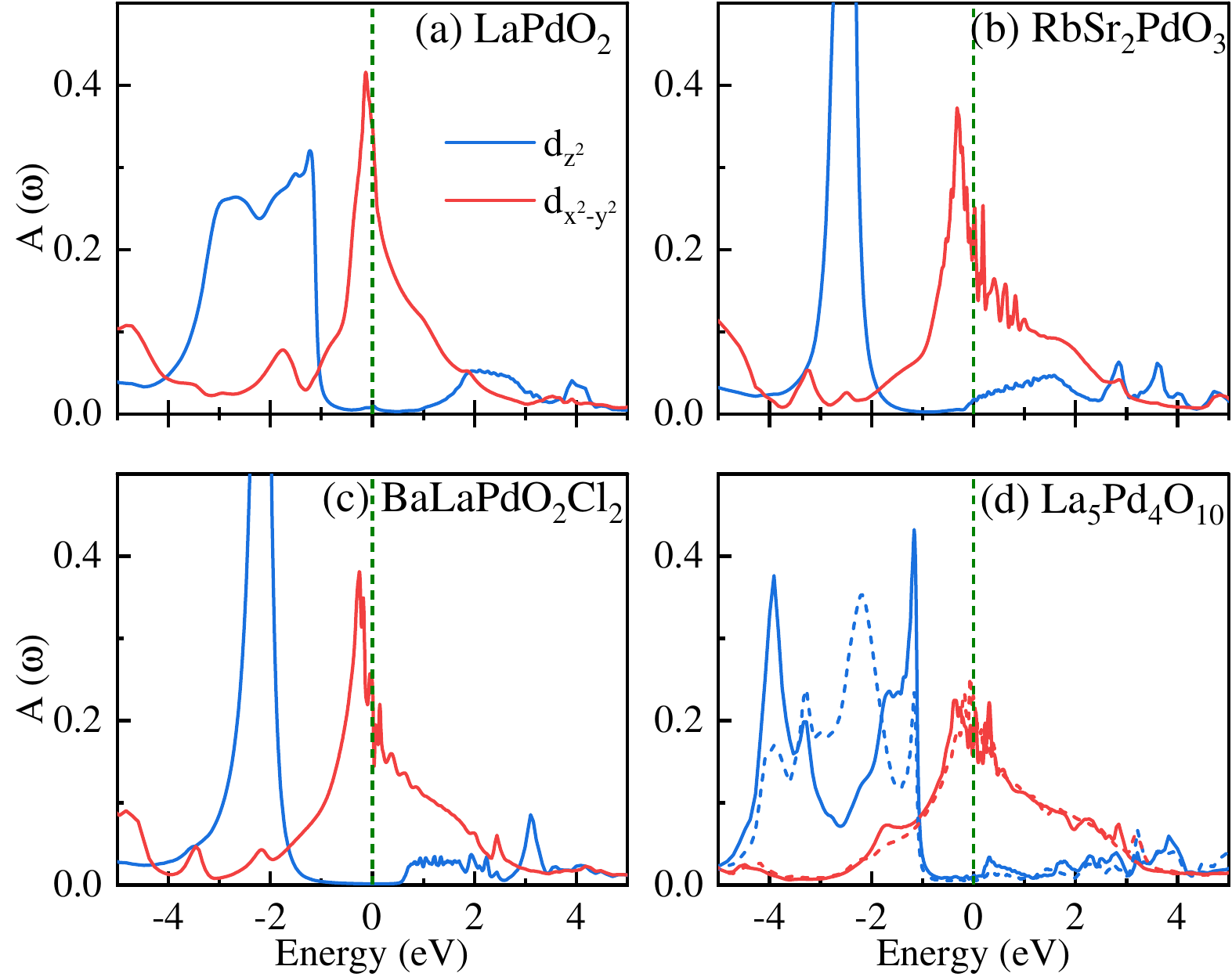}
  \caption{Orbital-resolved local spectral functions $A(\omega)$ obtained from DFT+DMFT at $T\approx290$ K for (a) LaPdO$_2$ at 0 GPa, (b) RbSr$_2$PdO$_3$ at 0 GPa, (c) BaLaPdO$_2$Cl$_2$ at 5 GPa, and (d) La$_5$Pd$_4$O$_{10}$ at 20 GPa. Blue and red curves denote the Pd-$d_{z^2}$ and Pd-$d_{x^2-y^2}$ orbitals, respectively. In panel (d), solid and dashed curves denote the inner- and outer-layer Pd sites, respectively. The vertical green dashed lines mark the Fermi level.}
  \label{fig:dos}
\end{figure}

\subsection{Local spectral functions}

Fig.~\ref{fig:dos} shows the local spectral functions $A(\omega)$ of the Pd $d_{z^2}$ and $d_{x^2-y^2}$ orbitals for the four layered palladates. We present La$_5$Pd$_4$O$_{10}$ as the representative multilayer compound in the spectral-function and band-structure plots, since La$_4$Pd$_3$O$_{8}$ exhibits similarly weak local correlations, as shown in Table~\ref{tab:occupancy} and Appendix C. In these compounds, the Pd-$d_{x^2-y^2}$ spectral weight is mainly concentrated near the Fermi level and remains finite at $\omega=0$, indicating that this orbital is the dominant low-energy channel within the Pd-$4d$ subspace. By contrast, the main occupied Pd-$d_{z^2}$ spectral peak lies below the Fermi level, with peak positions of approximately $-1.2$ eV in LaPdO$_2$, $-2.5$ eV in RbSr$_2$PdO$_3$, $-2.2$ eV in BaLaPdO$_2$Cl$_2$, and approximately $-1.1$ eV for the inner-layer Pd site
in La$_5$Pd$_4$O$_{10}$.

In cuprates, a larger energy separation between the Cu-$d_{x^2-y^2}$ and Cu-$d_{z^2}$ states reduces $d_{z^2}$ admixture into the low-energy band and favors a single-band description~\cite{sakakibara2012PRB}. In RbSr$_2$PdO$_3$ and BaLaPdO$_2$Cl$_2$, the main occupied Pd-$d_{z^2}$ peaks lie farther below $E_F$ than the corresponding peaks in LaPdO$_2$ and at the inner-layer Pd site in La$_5$Pd$_4$O$_{10}$, indicating a larger separation from the low-energy Pd-$d_{x^2-y^2}$ spectral weight. However, RbSr$_2$PdO$_3$ retains a small but finite Pd-$d_{z^2}$ contribution at $E_F$. The band and Fermi-surface results discussed below show that an additional electron pocket carries Pd-$d_{z^2}$ weight. This pocket may provide an additional self-doping channel and contribute to the smaller Pd-$d_{x^2-y^2}$ mass enhancement in RbSr$_2$PdO$_3$ compared with LaPdO$_2$ and BaLaPdO$_2$Cl$_2$. By contrast, BaLaPdO$_2$Cl$_2$ shows no discernible Pd-$d_{z^2}$ spectral weight at $E_F$ and has a predominantly Pd-$d_{x^2-y^2}$ local low-energy spectrum.

The difference in the Pd-$d_{z^2}$ spectral weight at $E_F$ may be related to the distinct spacer-layer environments. In RbSr$_2$PdO$_3$ and BaLaPdO$_2$Cl$_2$, adjacent PdO$_2$ planes are separated by different spacer layers, which can modify the electrostatic potential and local crystal field environment at the Pd sites. In particular, the Cl-containing spacer layers in BaLaPdO$_2$Cl$_2$ may help account for its negligible Pd-$d_{z^2}$ spectral weight at $E_F$.


For La$_5$Pd$_4$O$_{10}$, the solid and dashed curves in Fig.~\ref{fig:dos}(d) denote the inner- and outer-layer Pd sites, respectively. The local spectral functions of the two Pd sites are very similar near $E_F$, indicating weak layer dependence of the local low-energy spectra. This is consistent with the similar local occupations for the inner and outer Pd sites listed in Table~\ref{tab1}. For comparison, DFT+DMFT calculations for the high-pressure Ruddlesden--Popper phase of La$_4$Ni$_3$O$_{10}$ show weaker correlations in the inner-layer Ni-$e_g$ orbitals than in the outer-layer orbitals~\cite{wang2024nonfermi}.

\begin{figure*}[htb]
  \centering
  \includegraphics[width=18cm]{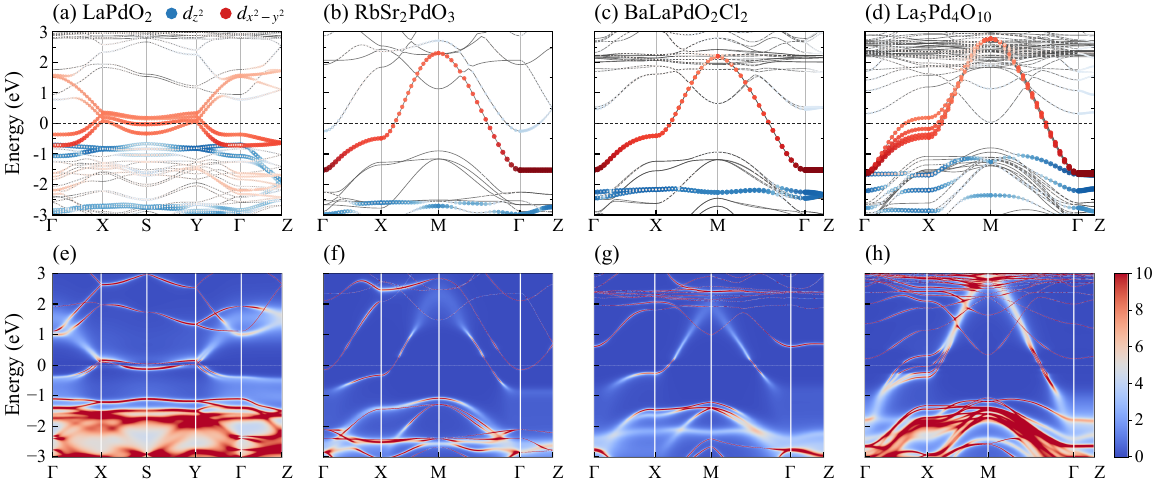}
  \caption{DFT band structures and DFT+DMFT momentum-resolved spectral functions of four layered palladates. Panels (a)--(d) present the DFT band structures with orbital projections, where the blue and red symbols denote the Pd-$d_{z^2}$ and Pd-$d_{x^2-y^2}$ orbitals, respectively. Panels (e)--(h) show the corresponding total momentum-resolved spectral functions $A(\mathbf{k},\omega)$ obtained from DFT+DMFT. The color scale represents the spectral intensity. The Fermi level is set to $E_F=0$ eV and marked by the horizontal dashed line. The four columns correspond to (a, e) LaPdO$_2$, (b, f) RbSr$_2$PdO$_3$, (c, g) BaLaPdO$_2$Cl$_2$, and (d, h) La$_5$Pd$_4$O$_{10}$, respectively. The structural optimization pressures are the same as those specified in Fig.~\ref{fig:self-energy}. The high-symmetry paths are chosen according to the crystal structure of each compound and therefore are not identical for all materials.}
  \label{fig:band-dft}
\end{figure*}

\begin{figure}[!h]
  \centering
  \includegraphics[width=8.6cm]{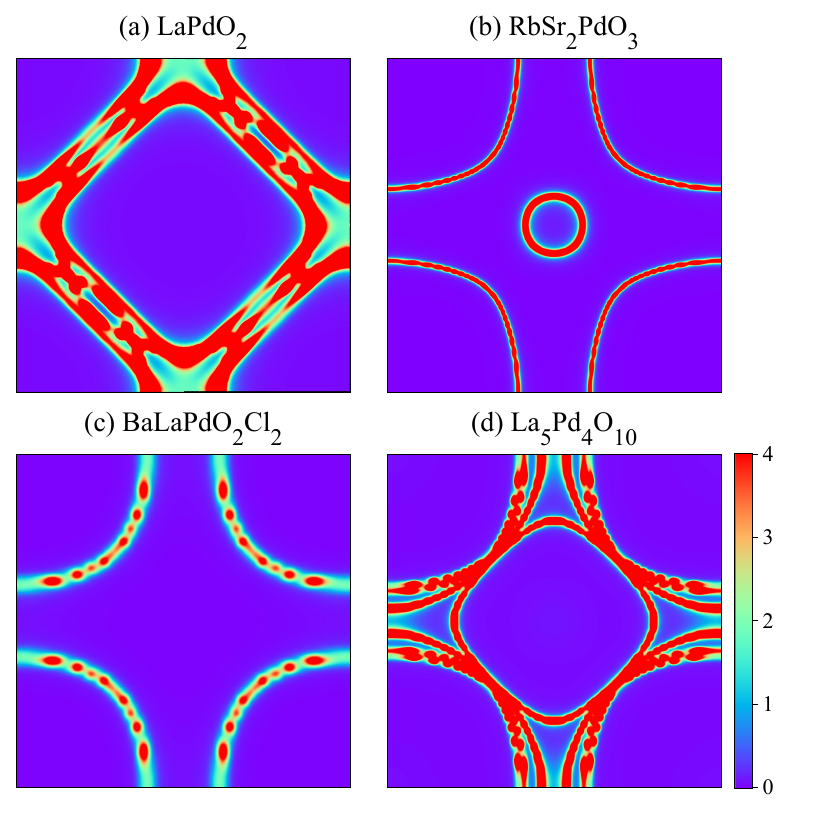}
  \caption{Fermi-level spectral-weight maps in the $k_z=0$ plane for selected layered palladates obtained by DFT+DMFT. Panels (a)--(d) show the momentum-resolved spectral intensity $A(\mathbf{k},\omega=0)$ for LaPdO$_2$ at 0 GPa, RbSr$_2$PdO$_3$ at 0 GPa, BaLaPdO$_2$Cl$_2$ at 5 GPa, and La$_5$Pd$_4$O$_{10}$ at 20 GPa, respectively. The color scale denotes the spectral intensity at the Fermi level.}
  \label{fig:fermi}
\end{figure}



\subsection{Correlated band structures}
We now discuss the orbital-projected DFT band structures and the DFT+DMFT momentum-resolved spectral functions $A(\mathbf{k},\omega)$ of the four layered palladates, shown in Fig.~\ref{fig:band-dft}. The DFT orbital projections show that the states near the Fermi level are mainly derived from the Pd-$e_g$ orbitals, with the low-energy electronic structure dominated by the in-plane Pd-$d_{x^2-y^2}$ orbital. Along the plotted high-symmetry paths, the Pd-$d_{x^2-y^2}$-dominated bands span approximately $2.5$ eV in LaPdO$_2$, about $3.5$ eV in RbSr$_2$PdO$_3$ and BaLaPdO$_2$Cl$_2$, and about $4.5$ eV in La$_5$Pd$_4$O$_{10}$. These differences indicate that the low-energy Pd-$d_{x^2-y^2}$ dispersion is sensitive to the structural environment, including the spacer-layer environment in RbSr$_2$PdO$_3$ and BaLaPdO$_2$Cl$_2$ and the interlayer coupling within the quadlayer PdO$_2$ block in La$_5$Pd$_4$O$_{10}$.

To further examine the effect of local electronic correlations on the low-energy electronic structure, Figs.~\ref{fig:band-dft}(e)--(h) show the DFT+DMFT momentum-resolved spectral functions $A(\mathbf{k},\omega)$ for the four layered palladates. Dispersive spectral features crossing $E_F$ remain visible in all four compounds, indicating metallic behavior at 290 K. Compared with the DFT bands, the low-energy Pd-$d_{x^2-y^2}$-derived dispersions are renormalized by local electronic correlations, while the main features of the band structures are retained.


Figure~\ref{fig:fermi} shows the total Fermi-level spectral weight $A(\mathbf{k},\omega=0)$ in the $k_z=0$ plane for the four layered palladates. LaPdO$_2$ exhibits multiple Fermi-level contours, as shown in Fig.~\ref{fig:fermi}(a). Its enlarged \textit{Pnma} unit cell introduces band folding. The orbital-projected DFT bands in Fig.~\ref{fig:band-dft}(a) nevertheless show that the states near $E_F$ are predominantly derived from the Pd-$d_{x^2-y^2}$ orbital. RbSr$_2$PdO$_3$ exhibits a large Fermi-surface contour and an additional electron pocket centered at $\Gamma$, as shown in Fig.~\ref{fig:fermi}(b). The DFT orbital projections in Fig.~\ref{fig:band-dft}(b) show that the corresponding electron band carries Pd-$d_{z^2}$ weight. This pocket persists in the DFT+DMFT spectra and may provide an self-doping channel. By contrast, BaLaPdO$_2$Cl$_2$ exhibits a single dominant contour without an additional pocket around $\Gamma$ in the $k_z=0$ plane shown in Fig.~\ref{fig:fermi}(c). Its low-energy bands are predominantly derived from Pd-$d_{x^2-y^2}$. For the structures studied here, BaLaPdO$_2$Cl$_2$ is closer to a cuprate-like single-band description than RbSr$_2$PdO$_3$.

In La$_5$Pd$_4$O$_{10}$, the four-layer PdO$_2$ block gives rise to multiple Pd-$d_{x^2-y^2}$-derived bands near $E_F$. The distinct Fermi-level contours in Fig.~\ref{fig:fermi}(d) are consistent with this
layer-derived band structure. Similar layer-derived bands occur in finite-layer square-planar nickelates $R_{n+1}$Ni$_n$O$_{2n+2}$~\cite{pan2022superconductivity,labollita2022correlated,worm2022correlations}.

\section{Discussion and Conclusion}

In summary, we have investigated the correlated electronic structures of layered palladates, LaPdO$_2$, RbSr$_2$PdO$_3$, BaLaPdO$_2$Cl$_2$, La$_4$Pd$_3$O$_8$, and La$_5$Pd$_4$O$_{10}$, using fully charge-self-consistent DFT+DMFT calculations. All these compounds contain square-planar PdO$_2$ layers and exhibit a low-energy electronic structure mainly derived from the Pd-$d_{x^2-y^2}$ orbital. However, their correlation strengths and degrees of single-band character are strongly material dependent.

LaPdO$_2$ and BaLaPdO$_2$Cl$_2$ show moderate quasiparticle renormalization in the Pd-$d_{x^2-y^2}$ channel, whereas RbSr$_2$PdO$_3$ and the multilayer compounds La$_4$Pd$_3$O$_8$ and La$_5$Pd$_4$O$_{10}$ are more weakly correlated and remain relatively coherent metals. This difference may reflect the combined influence of bandwidth, local crystal-field splitting, spacer-layer environment, and the presence or absence of self-doping channels near the Fermi level. Compared with the Cu-$3d$ orbitals in cuprates, the Pd-$4d$ orbitals are more spatially extended, which generally favors broader bands. Although these palladates exhibit a cuprate-like low-energy orbital character dominated by Pd-$d_{x^2-y^2}$, the present DFT+DMFT calculations yield weak to moderate mass renormalization and metallic spectra at 290~K.


For the optimized 5 GPa structure considered here, BaLaPdO$_2$Cl$_2$ realizes the cleanest Pd-$d_{x^2-y^2}$-dominated low-energy electronic structure among the compounds studied. The Pd-$d_{z^2}$ spectral weight is pushed away from the Fermi level, and a single dominant Fermi-level contour is found in the calculated $k_z=0$ plane. Together with its moderate mass enhancement, this makes BaLaPdO$_2$Cl$_2$ a promising palladate candidate for approaching a cuprate-like single-band regime. LaPdO$_2$ is less ideal in the relaxed bulk-like structure because of its lower symmetry, but its relatively stronger mass renormalization suggests that it remains an important reference system for tuning palladate correlations.

The present results also suggest several routes for future material design. First, LaPdO$_2$ may become more favorable if it can be stabilized as a high-symmetry thin film on a suitable substrate. Epitaxial strain could suppress the tilting distortion of the PdO$_4$ units, modify the Pd-$d_{x^2-y^2}$--Pd-$d_{z^2}$ level separation, and tune the Fermi-surface topology. In this sense, substrate-stabilized LaPdO$_2$ provides an interesting route toward a more cuprate-like palladate electronic structure.

Second, RbSr$_2$PdO$_3$ illustrates the importance of spacer-layer engineering. Although its low-energy bands are largely Pd-$d_{x^2-y^2}$-derived, the additional Fermi-level pocket indicates that the electronic structure is not completely single-band-like. This pocket is associated with interlayer-state and Pd-$d_{z^2}$ admixture and can be viewed as a self-doping channel that may weaken the effective correlation strength of the active Pd-$d_{x^2-y^2}$ band. Replacing Rb by other isovalent spacer ions, such as K, may shift the interlayer-derived states away from the Fermi level and reduce the unwanted Pd-$d_{z^2}$ self-doping. Such chemical control of the spacer layer could therefore be an effective way to optimize the single-band character of this family.

Finally, the multilayer palladates La$_4$Pd$_3$O$_8$ and La$_5$Pd$_4$O$_{10}$ provide a finite-layer route analogous to multilayer nickelates. Their low-energy states remain locally dominated by Pd-$d_{x^2-y^2}$ orbitals, but the layer degree of freedom generates multiple Fermi-surface sheets, while the calculated metallic states are relatively weakly correlated. This indicates that increasing the number of PdO$_2$ layers is not sufficient by itself to enhance correlations; instead, the optimal design should balance dimensionality, filling, bandwidth, and suppression of parasitic interlayer or Pd-$d_{z^2}$ states.


\begin{acknowledgments}
This work was supported by the National Key R\&D Program of China (Grants No. 2024YFA1408601 and No. 2024YFA1408602) and the National Natural Science Foundation of China (Grant No. 12434009). J.X.W. was also supported by the Outstanding Innovative Talents Cultivation Funded Programs 2025 of Renmin University of China. Z.Y.L. was also supported by the Innovation Program for Quantum Science and Technology (Grant No. 2021ZD0302402). Computational resources were provided by the Physical Laboratory of High Performance Computing in Renmin University of China.
\end{acknowledgments}

\appendix
\section{Details of the method}
\label{appA:method}

For the VASP calculations, the plane-wave energy cutoff was set to 550 eV. The electronic self-consistency convergence threshold was set to $10^{-8}$ eV. During structural optimization, both the lattice parameters and internal atomic coordinates were fully relaxed until the residual force on each atom was smaller than $0.01~\mathrm{eV}/\text{\AA}$. Phonon dispersions were calculated with PHONOPY using the finite-displacement method~\cite{Togo2023JPCM,Togo2023JPSJ}.

For the WIEN2k calculations, the muffin-tin radii $R_{\mathrm{MT}}$ for LaPdO$_2$, RbSr$_2$PdO$_3$, and BaLaPdO$_2$Cl$_2$ were chosen, when applicable, as 2.45, 2.50, 2.50, 2.20, 2.20, 1.90, and 2.50 bohr for La, Ba, Rb, Sr, Pd, O, and Cl, respectively. For La$_4$Pd$_3$O$_8$ and La$_5$Pd$_4$O$_{10}$, the corresponding radii were 2.35, 2.04, and 1.76 bohr for La, Pd, and O, respectively. The basis-set cutoff was set to $R_{\mathrm{MT}}K_{\max}=7.0$. The Brillouin zone was sampled using 1000 $k$ points for the charge-self-consistent DFT+DMFT calculations, while a denser mesh containing 20000 $k$ points was used for the calculation of the spectral functions.

The correlated subspace was constructed by projecting the Kohn--Sham states onto localized Pd-$4d$ orbitals within an energy window from $-10$ to $10$ eV relative to the Fermi level. The number of Monte Carlo steps was set to $10^8$ for each DMFT iteration. The DFT charge density, impurity self-energy, local Green's function, and chemical potential were updated self-consistently until convergence was reached.

\section{Optimized crystal structures and phonon stability}
\label{app:structure}

\begin{figure*}[htb]
  \centering
  \includegraphics[width=17.8cm]{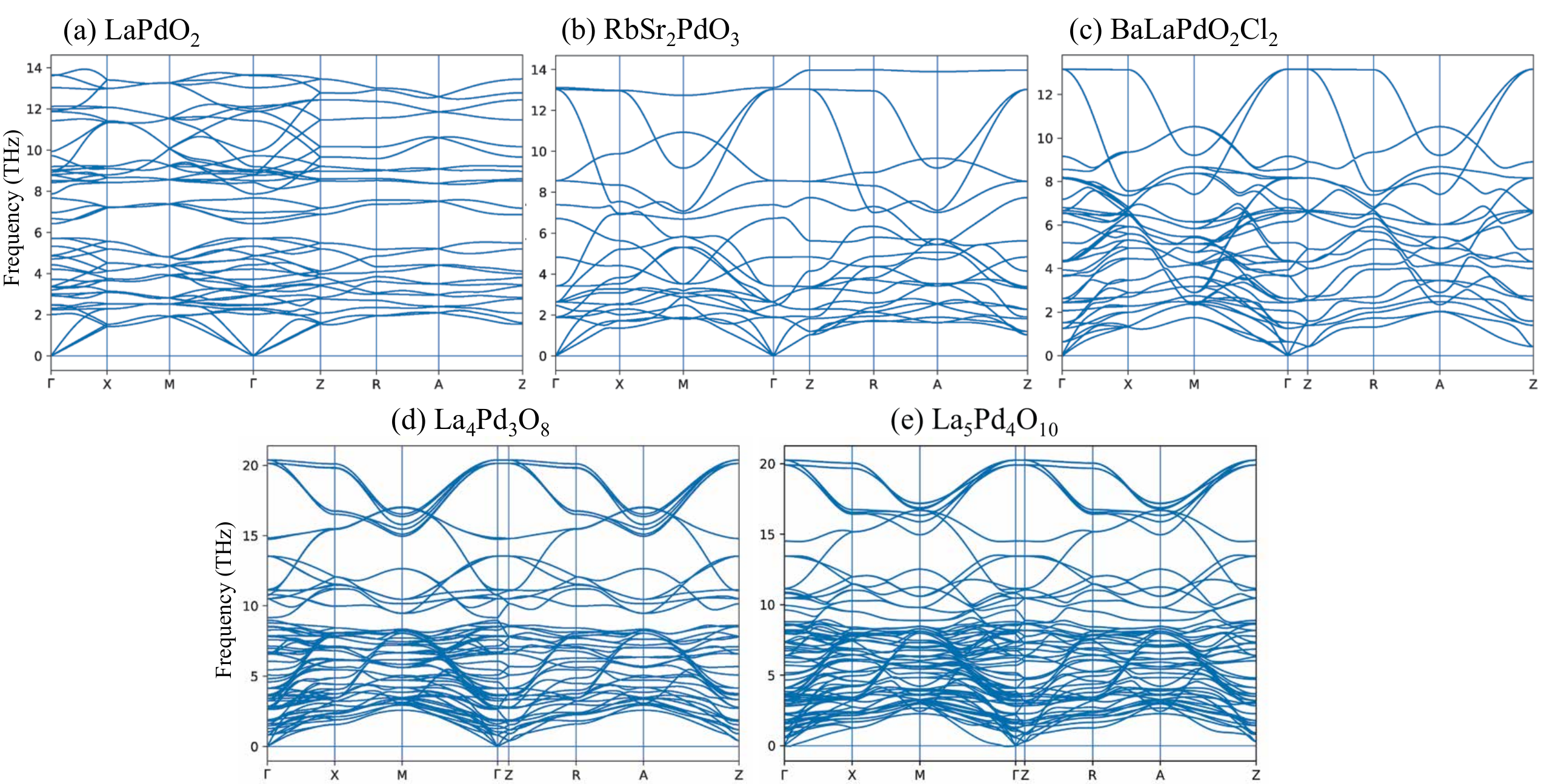}
  \caption{DFT-calculated phonon dispersions of the optimized structures of (a) LaPdO$_2$, (b) RbSr$_2$PdO$_3$, (c) BaLaPdO$_2$Cl$_2$, (d) La$_4$Pd$_3$O$_8$, and (e) La$_5$Pd$_4$O$_{10}$. The phonon frequencies are plotted along the high-symmetry paths of the corresponding Brillouin zones. The calculations were performed for the ambient-pressure optimized structures of LaPdO$_2$ and RbSr$_2$PdO$_3$, for the optimized 5 GPa structure of BaLaPdO$_2$Cl$_2$, and for the optimized 20 GPa structures of La$_4$Pd$_3$O$_8$ and La$_5$Pd$_4$O$_{10}$. The absence of imaginary phonon frequencies along the calculated high-symmetry paths supports the dynamical stability of these optimized structures within the present phonon calculations.}
  \label{fig:phonopy}
\end{figure*}

\begin{table}[t]
  \centering
  \renewcommand\arraystretch{1.2}
  \caption{DFT-optimized lattice parameters and space groups of the layered palladates at pressure $P$. The starting structural models follow previous theoretical studies~\cite{botana2018prm,Kitatani2023prl}, while the lattice constants listed here are obtained after structural optimization in this work. For all structures, $\alpha=\beta=\gamma=90^\circ$.}
  \label{tabA1}
  \setlength{\tabcolsep}{3pt}
  \begin{tabular*}{\columnwidth}{@{\extracolsep{\fill}}lccccc@{}}
    \hline\hline
    Material & $P$ (GPa) & Space group & $a$ (\AA) & $b$ (\AA) & $c$ (\AA) \\
    \hline
    LaPdO$_2$ & 0 & Pnma & 6.258 & 6.947 & 5.836 \\
    RbSr$_2$PdO$_3$ & 0 & P4/mmm & 4.342 & 4.342 & 7.653 \\
    BaLaPdO$_2$Cl$_2$ & 5 & I4mm & 4.097 & 4.097 & 12.633 \\
    La$_4$Pd$_3$O$_8$ & 0 & I4/mmm & 4.159 & 4.159 & 25.868 \\
    La$_5$Pd$_4$O$_{10}$ & 0 & I4/mmm & 4.170 & 4.170 & 32.528 \\
    La$_4$Pd$_3$O$_8$ & 20 & I4/mmm & 4.045 & 4.045 & 24.118 \\
    La$_5$Pd$_4$O$_{10}$ & 20 & I4/mmm & 4.056 & 4.056 & 30.327 \\
    \hline\hline
  \end{tabular*}
\end{table}

Table~\ref{tabA1} summarizes the DFT-optimized structural parameters used in the present DFT+DMFT calculations. The initial structural models were taken from previous theoretical studies of layered palladates, and all structures were subsequently relaxed within DFT. LaPdO$_2$ and RbSr$_2$PdO$_3$ were optimized at 0 GPa, and BaLaPdO$_2$Cl$_2$ was optimized at 5 GPa. For La$_4$Pd$_3$O$_8$ and La$_5$Pd$_4$O$_{10}$, structures were optimized at both 0 and 20 GPa. The 20 GPa structures were used for the main-text calculations. The optimized lattice constants and space groups are listed in Table~II. All optimized structures are orthorhombic or tetragonal, with $\alpha=\beta=\gamma=90^\circ$.

To examine the dynamical stability of the optimized structures, we calculated the phonon dispersions shown in Fig.~\ref{fig:phonopy}. For all five compounds, no imaginary phonon frequencies are found along the calculated high-symmetry paths in the Brillouin zone. The absence of imaginary phonon frequencies along these paths supports the dynamical stability of the structures at their respective optimization pressures within the present phonon calculations.

\section{\label{appC:pressure}Pd-$d_{x^2-y^2}$ self-energy in trilayer and quadlayer palladates under pressure}

\begin{figure}[htb]
  \centering
  \includegraphics[width=\columnwidth]{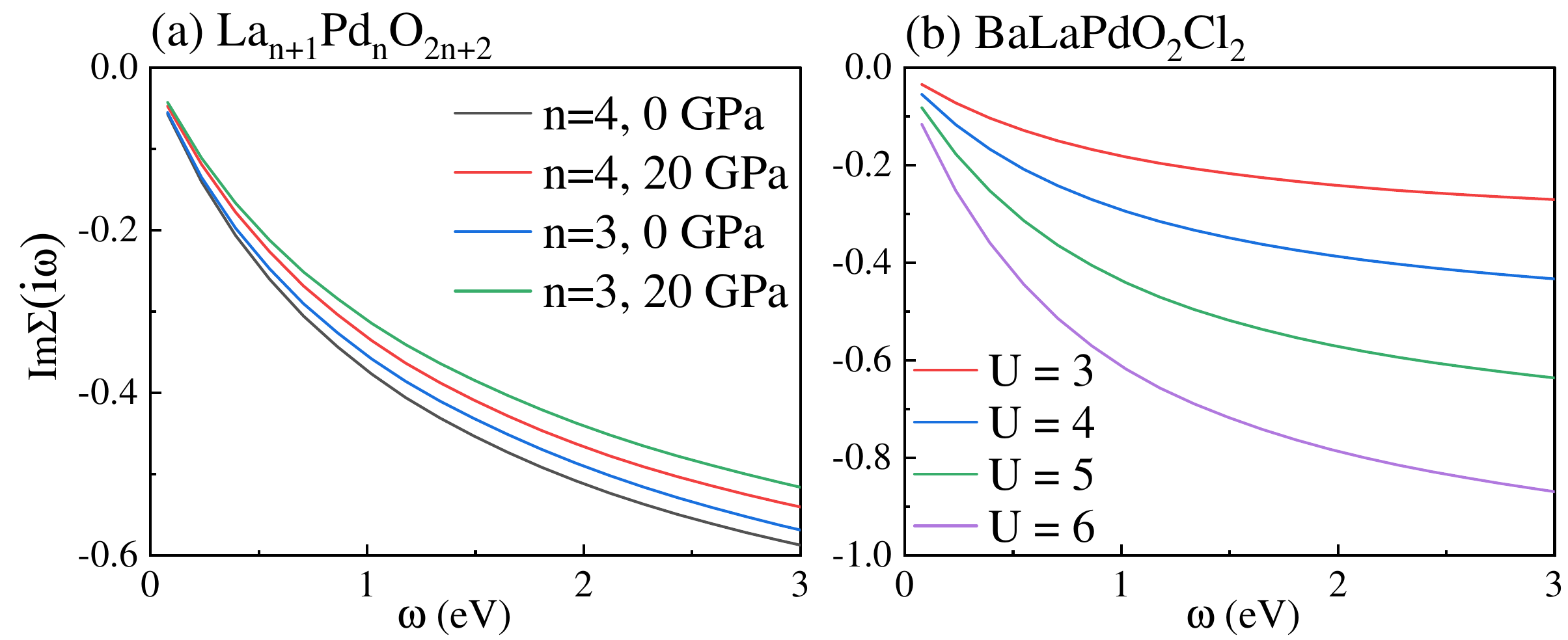}
  \caption{Pd-$d_{x^2-y^2}$ Matsubara self-energies, $\mathrm{Im}\Sigma(i\omega_n)$. (a) La$_4$Pd$_3$O$_8$ and La$_5$Pd$_4$O$_{10}$ at 0 and 20 GPa, averaged over inner- and outer-layer Pd sites. (b) BaLaPdO$_2$Cl$_2$ for $U=3$--6 eV at fixed $J_H=1~\mathrm{eV}$.}
  \label{fig:self-A2-A3}
\end{figure}

\begin{table}[t]
    \renewcommand\arraystretch{1.5}
    \caption{Local Pd-$4d$ orbital occupations and Pd-$d_{x^2-y^2}$ mass enhancement of BaLaPdO$_2$Cl$_2$ obtained from fully charge-self-consistent DFT+DMFT calculations for different values of $U$ at fixed $J_H=1~\mathrm{eV}$. The calculations are performed in the full five-orbital Pd-$4d$ correlated subspace. The values listed in the $n_{xz/yz}$ column denote the occupation of each of the two degenerate orbitals.}
    \label{tab:U_sensitivity}
    \begin{ruledtabular}
    \begin{tabular}{ccccccc}
    $U$ (eV) & $n_{z^2}$ & $n_{x^2-y^2}$ & $n_{xz/yz}$ & $n_{xy}$ & $N_{4d}$ & $Z^{-1}_{x^2-y^2}$ \\
    \hline
    3 & 1.76 & 1.39 & 1.92 & 1.94 & 8.93 & 1.40 \\
    4 & 1.74 & 1.36 & 1.91 & 1.93 & 8.85 & 1.66 \\
    5 & 1.73 & 1.32 & 1.91 & 1.93 & 8.80 & 1.71 \\
    6 & 1.71 & 1.28 & 1.91 & 1.93 & 8.74 & 2.28 \\
    \end{tabular}
    \end{ruledtabular}
\end{table}


Figure~\ref{fig:self-A2-A3}(a) compares the imaginary part of the Matsubara self-energy, $\mathrm{Im}\Sigma(i\omega_n)$, for the Pd-$d_{x^2-y^2}$ orbital in La$_4$Pd$_3$O$_8$ and La$_5$Pd$_4$O$_{10}$ at 0 and 20 GPa. For both compounds, applying pressure reduces the magnitude of $\mathrm{Im}\Sigma(i\omega_n)$ over the entire frequency range shown. The magnitude of the low-frequency slope also decreases under pressure, indicating weaker quasiparticle mass renormalization
in the Pd-$d_{x^2-y^2}$ orbital. This behavior is consistent with pressure-enhanced electronic itinerancy, which increases the effective bandwidth and drives the system toward a more coherent metallic regime.

At a fixed pressure, La$_5$Pd$_4$O$_{10}$ exhibits a slightly larger magnitude of the Pd-$d_{x^2-y^2}$ self-energy than La$_4$Pd$_3$O$_8$. The average nominal Pd valences are +1.25 in La$_5$Pd$_4$O$_{10}$ and +1.33 in La$_4$Pd$_3$O$_8$, corresponding to 0.25 and 0.33 holes per Pd relative to the $4d^9$ configuration, respectively.
Consistent with its lower nominal hole concentration, La$_5$Pd$_4$O$_{10}$ has higher Pd-$d_{x^2-y^2}$ occupations (1.17--1.21) than La$_4$Pd$_3$O$_8$ (1.14--1.16) at 20 GPa. However, their Pd-$d_{x^2-y^2}$ mass enhancements at this pressure remain nearly equal at $Z^{-1}=1.53$--1.54 and 1.50--1.51, respectively.

\section{\label{appD:U} Sensitivity to the Hubbard interaction \(U\) at fixed \(J_H\)}

To assess the sensitivity to the interaction strength, we performed additional Pd-$4d$ calculations for BaLaPdO$_2$Cl$_2$ by varying $U$ at fixed $J_H=1$ eV. Figure~\ref{fig:self-A2-A3}(b) shows the imaginary part of the Matsubara self-energy, $\mathrm{Im}\Sigma(i\omega_n)$, for the Pd-$d_{x^2-y^2}$ orbital. As expected, increasing $U$ systematically enhances the magnitude of $-\mathrm{Im}\Sigma(i\omega_n)$ and its low-frequency slope, indicating stronger quasiparticle renormalization. This trend is quantified in Table~\ref{tab:U_sensitivity}, where the mass enhancement increases from $Z^{-1}_{x^2-y^2}=1.40$ at $U=3$ eV to $2.28$ at $U=6$ eV. Thus, the numerical value of the mass enhancement is clearly interaction dependent.

However, the qualitative electronic structure remains unchanged over the tested range of $U$. The self-energy remains smooth at low Matsubara frequencies and does not show a divergent behavior characteristic of a correlation-induced insulating state. The Pd-$d_{x^2-y^2}$ orbital remains the least occupied and most strongly renormalized Pd-$4d$ orbital, whereas the Pd-$d_{z^2}$ and $t_{2g}$ orbitals stay close to full occupation. The total Pd-$4d$ projected occupation decreases only moderately, from $N_{4d}=8.93$ at $U=3$ eV to $N_{4d}=8.74$ at $U=6$ eV, reflecting a gradual enhancement of orbital polarization rather than a change of the low-energy orbital hierarchy. Thus, the hierarchy of local Pd-$4d$ orbital occupations in BaLaPdO$_2$Cl$_2$ is preserved over the tested range of $U=3$--6 eV at fixed $J_H=1$ eV, while the Pd-$d_{x^2-y^2}$ mass enhancement increases with $U$.


\bibliography{lco}

\end{document}